\documentclass[aps,pre,preprint,longbibliography]{revtex4-2}
\pdfoutput=1

\usepackage{amsmath,amssymb,amsfonts,graphicx,amsfonts}
\usepackage{textcomp,mathcomp}
\usepackage{color}
\usepackage[]{hyperref}
\usepackage{ascmac}
\usepackage{enumitem}
\usepackage{comment} 
\usepackage[skip=20pt]{subcaption}
\usepackage[braket, qm]{qcircuit}
\renewcommand{\vec}{\mathbf}

\usepackage{braket}
\usepackage{caption}
\usepackage{xcolor}

\allowdisplaybreaks[4]

\begin{document}
\begin{center}

{\Large Two-local modifications of SYK model with quantum chaos}

\end{center}
\vspace{0.1cm}
\vspace{0.1cm}
\begin{center}

Masanori Hanada,$^1$ Sam van Leuven,$^2$ Onur Oktay,$^3$ Masaki Tezuka$^4$

\end{center}
\vspace{0.3cm}

\begin{center}
$^1$School of Mathematical Sciences, Queen Mary University of London\\
Mile End Road, London, E1 4NS, United Kingdom\\
\vspace{1mm}
$^2$Mandelstam Institute for Theoretical Physics, School of Physics, National Institute for Theoretical and Computational Sciences (NITheCS), Gauteng, South Africa, and DSI-NRF Centre of Excellence in Mathematical and Statistical Sciences (CoE-MaSS),
University of the Witwatersrand, Johannesburg 2050, South Africa
\\
\vspace{1mm}
$^3$Department of Engineering Sciences, Abdullah Gül University, Kayseri 38080, Türkiye
\\
\vspace{1mm}
$^4$Department of Physics, Kyoto University\\
Kitashirakawa, Sakyo-ku, Kyoto 606-8502, Japan\\

\end{center}

\vspace{1.5cm}

\begin{center}
  {\bf Abstract}
\end{center}

The Sachdev--Ye--Kitaev (SYK) model may provide us with a good starting point for the experimental study of quantum chaos and holography in the laboratory. Still, the four-local interaction of fermions makes quantum simulation challenging, and it would be good to search for simpler models that keep the essence.
In this paper, we argue that the four-local interaction may not be important by introducing a few models that have two-local interactions. The first model is a generalization of the spin-SYK model, which is obtained by replacing the spin variables with SU($d$) generators. Simulations of this class of models might be straightforward on qudit-based quantum devices. We study the case of $d=3, 4, 5, 6$ numerically and observe quantum chaos already for two-local interactions in a wide energy range.
We also introduce modifications of spin-SYK and SYK models that have similar structures to the SU($d$) model (e.g., $H=\sum_{p,q}J_{pq}\chi_p\chi_{p+1}\chi_q\chi_{q+1}$ instead of the original SYK Hamiltonian $H=\sum_{p,q,r,s}J_{pqrs}\chi_p\chi_q\chi_r\chi_{s}$), which shows strongly chaotic features although the interaction is essentially two-local. 
These models may be a good starting point for the quantum simulation of the original SYK model.

\newpage

%\tableofcontents
%%%%%%%%%%%%%%%%%%%%%%
%%%%%%%%%%%%%%%%%%%%%%
\section{Introduction}
%%%%%%%%%%%%%%%%%%%%%%
%%%%%%%%%%%%%%%%%%%%%%
In this paper, we introduce a few models related to the SYK model~\cite{Sachdev:1992fk,Sachdev:2015efa,Maldacena:2016hyu} and study their properties. Our motivations are to understand quantum chaos better and to find models that can be simulated on quantum computers more easily. Ultimately, we want to find a good model of quantum chaos that can be simulated on quantum computers. A strong motivation comes from the experimental study of quantum gravity via holography~\cite{Danshita:2016xbo,Garcia-Alverez:2017,Pikulin:2017,Chew:2017,Chen:2018,Babbush:2019,Luo:2019,Wei:2021,Jafferis:2022crx,Asaduzzaman:2024}, i.e., we can simulate quantum gravitational systems if we can realize the dual quantum systems on quantum devices. 

In the context of quantum chaos, we would like to understand the essence underlying chaos in the SYK model. All-to-all interaction is not important, as we learned from the sparse SYK model~\cite{Xu:2020shn,Orman:2024mpw}. The use of fermions may not be important either, because the model of randomly coupled Pauli spins, known also as the spin-SYK model~\cite{Hanada:2023rkf,Swingle:2023nvv,Berkooz_2018}, is also strongly chaotic. A natural question then is whether the four-local nature of the interaction is essential. Certainly, chaos is not lost in $q$-local models with larger values of $q$. But what about $q=2$? At $q=2$, neither SYK nor spin-SYK is strongly chaotic. We show, however, that we can design models with two-local interactions which are strongly chaotic. Furthermore, some of the models we study can be regarded as special cases of the SYK model and, at the same time, special cases of the spin-SYK model. 

If we can simplify a model by keeping the essence, a potential bonus is that we might be able to study a simplified model on quantum computers. It could be a particularly useful approach until very powerful fault-tolerant quantum devices become available. In the future, quantum simulations could open up new directions in experimental studies of physics. A particularly interesting topic is the experimental study of quantum gravity via holography. Among the holographic approaches to quantum gravity, the SYK model is more tractable than the matrix model or Yang-Mills theory. However, it is still hard to put the original SYK model on a quantum computer, and significant efforts are invested to simplify the SYK model without losing the essence, specifically its strongly chaotic nature \cite{Xu:2020shn,Garcia-Garcia:2020cdo,Caceres_2021,Tezuka_2022,Anegawa_2023,Orman:2024mpw,Jha:2024vcw}. The spin-SYK model is one such attempt. The spin-SYK model can more easily be simulated on digital quantum computers because the Hamiltonian is simpler in terms of operators acting on qubits. Specifically, long Pauli strings appear when fermions are expressed in terms of Pauli operators, leading
to many two-qubit gates that are costly on NISQ devices.
 The spin-SYK model Hamiltonian does not contain these long Pauli strings. 

In this paper, we start with introducing a generalization of the spin-SYK model that may be realized naturally on a quantum device that uses qudits. Specifically, we replace spin operators, which are SU(2) generators in the fundamental representation by definition, with SU($d$) generators in the fundamental representation, e.g., for $d=3$, Gell-Mann matrices acting on qutrits. We call this model \textit{qudit SYK model}. Our numerical experiments show that these models are already chaotic for $q=2$. 
We also introduce modifications of SYK and spin-SYK models that have natural connections to the qudit SYK model. These models are defined in Sec.~\ref{sec:models}. Among them, the overlapping clusters SYK model (Sec.~\ref{sec:q2M2SYK}) has a particularly simple form. Furthermore, by taking the parameter $M$ (cluster size) large, the original SYK model is restored. The existence of a systematic way to approach the original SYK model may be useful in establishing the signatures of quantum gravity in quantum simulations. In this context, it may be worth mentioning a recent work~\cite{Jafferis:2022crx} that claimed the experimental realization of the wormhole. This interesting paper was challenged by ref.~\cite{Kobrin:2023rzr} (see also \cite{Orman:2024mpw}). The problem at the center of the debate is whether the signal observed in the simulations can survive while a simplified Hamiltonian used in ref.~\cite{Jafferis:2022crx} is deformed gradually to the original SYK Hamiltonian. Phrased differently, the problem is whether there is a systematic way to simplify the SYK model so that the Hamiltonian used in ref.~\cite{Jafferis:2022crx} is obtained. Perhaps, by simulating the overlapping clusters SYK on quantum devices, stronger evidence can be provided. 

This paper is organized as follows. 
Sec.~\ref{sec:models} provides the definitions of the models we study in this paper, and the results of our numerical analyses are shown in Sec.~\ref{sec:Numerics}. 
The qudit SYK model is defined in Sec.~\ref{sec:qudit_model_definition} and numerically studied in Sec.~\ref{sec:Numerics-qudit}. 
The clusters spin-SYK model and clusters SYK model are defined in Sec.~\ref{sec:clusters-spin-SYK} and Sec.~\ref{sec:clusters-SYK}, respectively. 
The overlapping clusters SYK model is studied numerically in Sec.~\ref{sec:Numerics-SYK}. 
Sec.~\ref{sec:conclusion} is devoted to conclusions and discussion. 
%%%%%%%%%%%%%%%%%%%%%%
%%%%%%%%%%%%%%%%%%%%%%
\section{Models}\label{sec:models}
%%%%%%%%%%%%%%%%%%%%%%
%%%%%%%%%%%%%%%%%%%%%%
In this section, we will introduce a few models. Specifically:
\begin{enumerate}
    \item 
    Qudit SYK model. This is obtained by replacing qubits and spin operators in the spin-SYK model with qudits and SU($d$) generators. 
    
    \item 
    Clusters spin-SYK model. This is obtained by restricting interaction terms in the spin-SYK model to the ones with a certain cluster property. It can also be obtained from the qudit SYK model with $d=2^M$ ($M=1,2,\ldots$) restricting the SU($d$) generators to a certain subclass. Here, $M$ denotes the number of qubits grouped to form one qudit.

    \item 
    Clusters SYK model. This is obtained by restricting interaction terms in the SYK model to the ones with a certain cluster property. This can also be obtained from the clusters spin-SYK model replacing the spin operators with fermions. 
\end{enumerate}
Below, we will explain these models. In Sec.~\ref{sec:Numerics}, we will study the qudit SYK model and clusters SYK model (specifically, overlapping clusters SYK model) numerically. 
%%%%%%%%%%%%%%%%%%%%%%
%%%%%%%%%%%%%%%%%%%%%%
\subsection{%Non-locally interacting qudits
Qudit SYK model}\label{sec:qudit_model_definition}
%%%%%%%%%%%%%%%%%%%%%%
%%%%%%%%%%%%%%%%%%%%%%
The system of $L$ qudits (see e.g., refs.~\cite{Wang2020,Ringbauer:2021lhi,Chi:2022uql}) provides a natural generalization of the system of qubits. Each qudit has $d$ degrees of freedom, i.e., $d=2$ for qubit, $d=3$ for qutrit, and $d=4$ for ququart. 
The dimension of the Hilbert space is $d^L$. 

Let $\tau_\alpha$ ($\alpha=1,\ldots,d^2-1$) be the generators of SU($d$) algebra in the fundamental representation normalized as ${\rm Tr}(\tau_\alpha\tau_\beta)=2\delta_{\alpha\beta}$.
For the SU(2) theory, an explicit example of the choice of such generators is the set of the Pauli matrices, 
\begin{align}
\sigma_1=X=\begin{pmatrix}0&1\\1&0\end{pmatrix}\, , 
\qquad
\sigma_2=Y=\begin{pmatrix}0&-\mathrm{i}\\\mathrm{i}&0\end{pmatrix}\, , 
\qquad
\sigma_3=Z=\begin{pmatrix}1&0\\0&-1\end{pmatrix}\, . 
\end{align}
For the SU($3$) theory, we can use the Gell-Mann matrices defined by 
\begin{align}
&
\tau_1
=
\left(
\begin{array}{ccc}
0 & 1 & 0\\
1 & 0 & 0\\
0 & 0 & 0
\end{array}
\right)\, , 
\quad
\tau_2
=
\left(
\begin{array}{ccc}
0 & -\mathrm{i} & 0\\
\mathrm{i} & 0 & 0\\
0 & 0 & 0
\end{array}
\right)\, , 
\quad
\tau_3
=
\left(
\begin{array}{ccc}
1 & 0 & 0\\
0 & -1 & 0\\
0 & 0 & 0
\end{array}
\right)\, , 
\nonumber\\
&
\tau_4
=
\left(
\begin{array}{ccc}
0 & 0 & 1\\
0 & 0 & 0\\
1 & 0 & 0
\end{array}
\right)\, , 
\quad
\tau_5
=
\left(
\begin{array}{ccc}
0 & 0 & -\mathrm{i}\\
0 & 0 & 0\\
\mathrm{i} & 0 & 0
\end{array}
\right)\, , 
\nonumber\\
&
\tau_6
=
\left(
\begin{array}{ccc}
0 & 0 & 0\\
0 & 0 & 1\\
0 & 1 & 0
\end{array}
\right)\, , 
\quad
\tau_7
=
\left(
\begin{array}{ccc}
0 & 0 & 0\\
0 & 0 & -\mathrm{i}\\
0 & \mathrm{i} & 0
\end{array}
\right)\, , 
\quad\tau_8
=
\frac{1}{\sqrt{3}}
\left(
\begin{array}{ccc}
1 & 0 & 0\\
0 & 1 & 0\\
0 & 0 & -2
\end{array}
\right)\, . 
\end{align}
For SU($d$) theory, we can use $S_{ab}$, $A_{ab}$ ($a<b$), and $D_n$ ($n=1,\ldots,d-1$), 
where
\begin{align}
(S_{ab})_{ij}\equiv\delta_{ai}\delta_{bj}+\delta_{aj}\delta_{bi}, 
\quad
(A_{ab})_{ij}\equiv\mathrm{i}\left(\delta_{ai}\delta_{bj}-\delta_{aj}\delta_{bi}\right)
\end{align}
and
\begin{align}
D_{n}\equiv\textrm{diag}(\underbrace{1,\ldots,1}_n,-n,0,\ldots,0)\times \sqrt{\frac{2}{n(n+1)}}\, .
\end{align}
We use the notation $T_{i,\alpha}$ to mean a generator $\tau_\alpha$ acting on the $i$-th qudit. For other qudits, it acts trivially as identity, $I_d$. Namely, 
\begin{align}
    T_{1,\alpha} &= \tau_\alpha\otimes I_d\otimes I_d\otimes\cdots\otimes I_d\, , 
    \nonumber\\
    T_{2,\alpha} &= I_d\otimes\tau_\alpha\otimes I_d\otimes\cdots\otimes I_d\, ,   
    \nonumber\\
    & \qquad\qquad\vdots
        \nonumber\\
    T_{L,\alpha} &= I_d\otimes I_d\otimes\cdots\otimes I_d\otimes\tau_\alpha\, . 
\end{align}

A natural generalization of spin-SYK model is given by the Hamiltonian consisting of random couplings of these SU($d$) generators,\footnote{
For $q=2$, this Hamiltonian reduces to the Sachdev-Ye model~\cite{Sachdev:1992fk} if the random couplings are taken independent of the adjoint indices as $J_{i_1,\alpha_1;i_2,\alpha_2}=J_{i_1i_2}\delta_{\alpha_1,\alpha_2}$.
} 
\begin{align}
    H
    =
    \sum_{i_1<i_2<\cdots<i_q}
    \sum_{\alpha_1,\ldots,\alpha_q}
    J_{i_1,\alpha_1;\ldots;i_q,\alpha_q}
    T_{i_1,\alpha_1}\cdots T_{i_q,\alpha_q}\, . \label{Hamiltonian:Qudit_SYK}
\end{align}
We take $J_{i_1,\alpha_1;\cdots;i_q,\alpha_q}$ Gaussian random with variance $J^2$. 
The number of terms is $\binom{L}{q}\times (d^2-1)^q$.

For $d=2$, this is the spin-SYK model. One of the motivations to study the spin-SYK model is that it is a potentially interesting model of quantum chaos simulatable on qubit-based quantum devices. For $d>2$, we may be able to find a good model of quantum chaos that can be simulated on qudit-based devices. In this context, a nontrivial question is: what is the smallest $q$ for the system to be chaotic for each $d$? In Sec.~\ref{sec:Numerics-qudit}, we study $d=3,4,5$ and $6$ numerically, and we find that $q=2$ is strongly chaotic. This property can make quantum simulation more tractable.
The large-$d$ limit with fixed $L$ may also be useful, though we do not study this case in this paper.

%%%%%%%%%%%%%%%%%%%%%%
%%%%%%%%%%%%%%%%%%%%%%
\subsection{Clusters models}\label{sec:clusters}
%%%%%%%%%%%%%%%%%%%%%%
%%%%%%%%%%%%%%%%%%%%%%
For the qudit SYK model introduced above, specific details of the operators acting on each qudit should not be important. As an alternative, we define a modification of spin-SYK model and SYK model that can be studied straightforwardly on qubit-based device (not qudit-based device). 
More specifically, in the following we define three closely related models, where the first two models, the clusters spin-SYK and the (gauged) clusters SYK serve as motivation for the definition of our final model, the overlapping clusters SYK model. The latter model turns out to be the simplest to formulate, it exhibits strongly chaotic features and has a parameter, the cluster size $M$, which interpolates the model between a two-local model and the full SYK$_4$ model.
Among these models, we will study the overlapping clusters SYK model numerically in Sec.~\ref{sec:Numerics-SYK}.

%%%%%%%%%%%%%%%%%%%%%%
%%%%%%%%%%%%%%%%%%%%%%
\subsubsection{Clusters spin-SYK model}\label{sec:clusters-spin-SYK}
%%%%%%%%%%%%%%%%%%%%%%
%%%%%%%%%%%%%%%%%%%%%%
To define the clusters spin-SYK model, we start with the $q=2$ SU(4) qudit SYK from Section \ref{sec:qudit_model_definition}. To realize this model using qubits, such that it can be studied more easily on qubit machines, we take a tensor product of two qubits to form a ququart and replace SU(4) generators with the tensor products of two Pauli operators. Specifically, we investigate the following Hamiltonian:
\begin{align}
    H
    =
    \sum_{i_1<i_2}
\sum_{\alpha_1,\beta_1,\alpha_2,\beta_2=1}^3
J_{i_1,\alpha_1,\beta_1;i_2,\alpha_2,\beta_2}\, 
    (\sigma_{2i_1-1,\alpha_1}
    \sigma_{2i_1,\beta_1})
    (\sigma_{2i_2-1,\alpha_2}
    \sigma_{2i_2,\beta_2})\, . 
    \label{Hamiltonian:cluster_spin_SYK}
\end{align}
Here, $\sigma_{j,\alpha}$ are Pauli operators belonging to the $j$-th qubit, and the pair of $(2i-1)$-th and $2i$-th qubits is identified with the $i$-th ququart. 
Note that the interactions in this model are two-local in the sense that two pairs of neighboring qubits interact with each other.
In what follows, “two-local” refers to two-cluster locality: each interaction term couples exactly two clusters (effective sites) and, when a cluster is realized by two qubits, the same term acts on four physical qubits.

The Hamiltonian above can be seen as a non-local interaction between two clusters consisting of six Pauli operators acting on two spins.  It is straightforward to generalize it to the $\tilde{q}$-local interaction between $\tilde{q}$ clusters consisting of more spins, where we use $\tilde{q}$ to distinguish from $q$, the total number of operators in the interaction terms of (spin-)SYK models. We will use $M$ to indicate the number of distinct operators in a given cluster, which quantifies the cluster size. We call these models the \textit{$\tilde{q}$-local $M$-clusters spin-SYK models}. 

%%%%%%%%%%%%%%%%%%%%%%
%%%%%%%%%%%%%%%%%%%%%%
\subsubsection{Gauged clusters SYK model}\label{sec:clusters-SYK}
%%%%%%%%%%%%%%%%%%%%%%
%%%%%%%%%%%%%%%%%%%%%%

Suppose we restrict Pauli operators in the two-local clusters spin-SYK model in \eqref{Hamiltonian:cluster_spin_SYK} to just two Pauli matrices $\sigma_1=X$ and $\sigma_2=Y$. This is a restricted version of spin-XY4 model~\cite{Hanada:2023rkf}, but it can also be regarded as a clustered version of the SYK model for the following reason.  

The fermions $\chi_j$ ($j=1,2,\ldots,N=4L$) in the SYK model, that satisfy the anticommutation relation $\{\chi_j,\chi_k\}=2\delta_{jk}$, can be expressed by Pauli operators as \footnote{Here, $X_j$, $Y_j$, $Z_j$ mean the $X$, $Y$, $Z$ gates acting on the $j$th qubit. Equivalently, $X_1=X\otimes I\otimes\cdots I$, $Y_2=I\times Y\otimes I\otimes\cdots I$, etc.}
\begin{align}
\chi_{1}
=
X_1\, , 
\qquad
\chi_{2}
=
Y_1,
\label{Jordan-Wigner-1}
\end{align}
and
\begin{align}
\chi_{2j-1}
=
Z_{1}
\cdots
Z_{j-1}
X_{j}\, ,
\qquad
\chi_{2j}
=
Z_{1}
\cdots
Z_{j-1}
Y_{j},
\label{Jordan-Wigner-2}
\end{align}
for $j\ge 2$. 
By introducing notation
\begin{align}
    \psi_{j,1}\equiv\chi_{2j-1}\, , 
    \qquad
    \psi_{j,2}\equiv\chi_{2j}\, , 
    \label{numbering_fermion}
\end{align}
we can relate the tensor products of Pauli operators and those of fermions as
\begin{align}
&
X_{j-1}
X_{j}
=
-\mathrm{i}%\cdot 
\psi_{j-1,2}
\psi_{j,1}\, ,
\qquad
X_{j-1}
Y_{j}
=
-\mathrm{i}%\cdot 
\psi_{j-1,2}
\psi_{j,2}\, ,
\nonumber\\
&
Y_{j-1}
X_{j}
=
\mathrm{i}%\cdot 
\psi_{j-1,1}
\psi_{j,1}\, ,
\qquad
Y_{j-1}
Y_{j}
=
\mathrm{i}%\cdot 
\psi_{j-1,1}
\psi_{j,2}\, . 
\end{align}
Therefore, when $\alpha_1,\beta_1,\alpha_2,\beta_2$ are restricted to 1 or 2 (equivalently, $x$ and $y$), \eqref{Hamiltonian:cluster_spin_SYK} can be rewritten as 
\begin{align}
    H
    =
    \sum_{i_1<i_2}
\sum_{\alpha_1,\beta_1,\alpha_2,\beta_2=1}^2
J'_{i_1,\alpha_1,\beta_1;i_2,\alpha_2,\beta_2}\, 
    (\psi_{2i_1-1,\alpha_1}
    \psi_{2i_1,\beta_1})
    (\psi_{2i_2-1,\alpha_2}
    \psi_{2i_2,\beta_2})\, . 
    \label{Hamiltonian:naive_cluster_SYK}
\end{align} 
This is a restricted version of the SYK model which is two-local in the sense explained in Section \ref{sec:clusters-spin-SYK}. 
The Hamiltonian can be seen as a non-local interaction between two clusters of fermions. In each interaction term, two fermions are chosen out of four fermions per cluster. It is straightforward to generalize this to the $\tilde{q}$-local interaction between $\tilde{q}$ clusters consisting of $M$ fermions. The example above is $\tilde{q}=2$ and $M=4$. In the non‑overlapping clusters SYK models, the cluster size $M$ denotes the number of Majorana fermions per cluster (equivalently, $M$/2 qubits).

The fact that we do not have to use fermions makes digital quantum simulation easier. Unfortunately, however, the $\tilde{q}=2$, $M=4$ version we defined above is too simple to be chaotic for the following reasons. 

This model has many conserved charges. Specifically, $Z_{j-1}Z_{j}$ commutes with the Hamiltonian for any $j$, and hence, there is conserved `parity' for each cluster, corresponding to the eigenvalues $\pm 1$ of $Z_{j-1}Z_{j}$.  Therefore, the Hilbert space splits into $2^L$ parity sectors and each of them is $2^L$-dimensional, where $L=N/4$ is the total number of clusters. To see if this system is chaotic, one should compare a fixed parity sector with random matrices. 

Let us assume all $L$ parities are $+1$. In Ref.~\cite{Swingle:2023nvv}, this restriction is introduced as a gauge constraint, and the model in this sector is called the \textit{gauged} clusters model. 

Then, for each $j$, the Hilbert space is spanned by $\ket{\uparrow\uparrow}$ and $\ket{\downarrow\downarrow}$, and 
\begin{align}
&
    X_{j-1}X_{j}
    \ket{\uparrow\uparrow}
    =
    \ket{\downarrow\downarrow}\, , 
    \qquad
    X_{j-1}X_{j}
    \ket{\downarrow\downarrow}
    =
    \ket{\uparrow\uparrow}\, , 
    \nonumber\\
&
    Y_{j-1}Y_{j}
    \ket{\uparrow\uparrow}
    =
    -\ket{\downarrow\downarrow}\, , 
    \qquad
    Y_{j-1}Y_{j}
    \ket{\downarrow\downarrow}
    =
    -\ket{\uparrow\uparrow}\, , 
    \nonumber\\    
    &
    X_{j-1}Y_{j}
    \ket{\uparrow\uparrow}
    =
    \textrm{i}\ket{\downarrow\downarrow}\, , 
    \qquad
    X_{j-1}Y_{j}
    \ket{\downarrow\downarrow}
    =
    -\textrm{i}\ket{\uparrow\uparrow}\, , 
    \nonumber\\
        &
    Y_{j-1}X_{j}
    \ket{\uparrow\uparrow}
    =
    \textrm{i}\ket{\downarrow\downarrow}\, , 
    \qquad
    Y_{j-1}X_{j}
    \ket{\downarrow\downarrow}
    =
    -\textrm{i}\ket{\uparrow\uparrow}\, .
\end{align}
This means that $X_{j-1}X_{j}$ and $Y_{j-1}Y_{j}$ act as Pauli-X on this 2-dimensional space, and  $X_{j-1}Y_{j}$ and $Y_{j-1}X_{j}$ act as Pauli-Y, identifying $\ket{\uparrow\uparrow}$ and $\ket{\downarrow\downarrow}$ as $\begin{pmatrix}1\\ 0\end{pmatrix}$ and $\begin{pmatrix}0\\ 1\end{pmatrix}$, respectively:
\begin{align}
        \left.
X_{j-1}X_{j}
    \right|_+
    =
    X\, , 
    \qquad
            \left.  
Y_{j-1}Y_{j}
    \right|_+
    =
    -X\, , 
\qquad
            \left.
X_{j-1}Y_{j}
    \right|_+
    =
    Y\, , 
    \qquad
            \left.  
Y_{j-1}X_{j}
    \right|_+
    =
    Y\, . 
\end{align}
Therefore, this sector reduces to the $q=2$ spin-XY model, which we do not expect to be maximally chaotic at low temperatures in the large-$N$ limit \cite{Swingle:2023nvv,basu2025complexityquadraticquantumchaos}. 
To overcome this problem, we make the model slightly more complicated, while keeping it essentially two-local. 
Below, we discuss a few options. 
In Sec.~\ref{sec:Numerics-SYK}, we will study the option introduced in Sec.~\ref{sec:q2M2SYK} numerically. 
%%%%%%%%%%%%%%%%%%%%%%
%%%%%%%%%%%%%%%%%%%%%%
\paragraph{Clusters SYK with $M>4$}

%%%%%%%%%%%%%%%%%%%%%%
%%%%%%%%%%%%%%%%%%%%%%
One of the simplest resolutions is to take the number of Majorana fermions $M$ in the cluster larger than four~\cite{Swingle:2023nvv}. 
(As before, we take two fermions from each cluster in the interaction terms.)
Let $M=6$ and $\mathcal{I}_j=\{3j-2,3j-1,3j\}$ for $j=1,2,\ldots,L$. 
We consider a Hamiltonian defined by 
\begin{align}
    H
    =
     \sum_{i<j}
    \sum_{r,r'\in \mathcal{I}_i}
    \sum_{s,s'\in \mathcal{I}_j}
    \sum_{\alpha,\alpha',\beta,\beta'=1}^2
    J_{rr'ss'\alpha\alpha'\beta\beta'}\, 
    \psi_{r,\alpha}
    \psi_{r',\alpha'}
    \psi_{s,\beta}
    \psi_{s',\beta'}\, 
\label{Hamiltonian:restricted_spin_SYK_ver_1}
\end{align}
The total number of Majorana fermions is $ML=6L$. 
We assume $r<r'$ and $s<s'$. Then, from $\mathcal{I}_j$, we have
\begin{align}
    \sigma_{\alpha,3j-2}\sigma_{\alpha',3j-1}\, ,
    \qquad
    \sigma_{\alpha,3j-1}\sigma_{\alpha',3j}
    \qquad
    \sigma_{\alpha,3j-2}Z_{3j-1}\sigma_{\alpha',3j}\, , 
\end{align}
where $\alpha$ and $\alpha'$ are $x$ or $y$. They commute with $\sigma_{z,3j-2}Z_{3j-1}\sigma_{z,3j}$. Therefore, $\pm 1$ `parity' is conserved. 
Let us take the parity-$+$ sector consisting of $\ket{\uparrow\uparrow\uparrow}$, $\ket{\uparrow\downarrow\downarrow}$ , $\ket{\downarrow\uparrow\downarrow}$  and $\ket{\downarrow\downarrow\uparrow}$. Relating them to a two-qubit basis $\begin{pmatrix}1\\ 0 \\ 0 \\ 0\end{pmatrix}$, $\begin{pmatrix}0\\ 1 \\ 0 \\ 0\end{pmatrix}$, 
$\begin{pmatrix}0\\ 0 \\ 1 \\ 0\end{pmatrix}$, and 
$\begin{pmatrix}0\\ 0 \\ 0 \\ 1\end{pmatrix}$ in this order, we can write these operators as 
\begin{align}
   \left.
    X_{3j-2}X_{3j-1}
    \right|_+
   =
    \left(
\begin{array}{cccc}
0  & 0 & 0 & 1\\
0  &  0 & 1 & 0 \\
0 & 1 & 0 & 0 \\
1 & 0  & 0 & 0    
\end{array}
    \right)
    =
    X\otimes X\, , \\
%\end{align}
%
%\begin{align}
    \left.
    X_{3j-1}X_{3j}
    \right|_+
    =
    \left(
\begin{array}{cccc}
 0 & 1& 0 & 0 \\
 1& 0 & 0 & 0 \\
 0 &  0 & 0 & 1\\
 0 & 0 & 1 & 0
\end{array}
    \right)
    =
    I\otimes X\, , \\
%\end{align}
%
%\begin{align}
    \left.
  X_{3j-2}Z_{3j-1}X_{3j}
  \right|_+
    =
    \left(
\begin{array}{cccc}
0 & 0 & 1 & 0 \\
0 & 0 & 0 & -1\\
 1 & 0  & 0  & 0 \\
 0 & -1 & 0 & 0
\end{array}
    \right)
    =
    X\otimes Z\, ,  
\end{align}
\begin{align}
%    \left.
%X_{3j-2}X_{3j-1}
%    \right|_+
%    &=
%    X\otimes X\, , 
%    \qquad
%\left.
%    X_{3j-1}X_{3j}
%    \right|_+
%    =
%    I\otimes X\, , 
%    \qquad\, \, \, 
%\left.
%    X_{3j-2}Z_{3j-1}X_{3j}
%    \right|_+
%    =
%    X\otimes Z\, , 
%    \nonumber\\
    %%%%%
    \left.
X_{3j-2}Y_{3j-1}
    \right|_+
    &=
    X\otimes Y\, , 
    \qquad\ 
\left.
    X_{3j-1}Y_{3j}
    \right|_+
    =
    Z\otimes Y\, , 
    \qquad\, \, \, \, 
\left.
    X_{3j-2}Z_{3j-1}Y_{3j}
    \right|_+
    =
    Y\otimes I\, , 
    \nonumber\\
    %%%%%
        \left.
Y_{3j-2}X_{3j-1}
    \right|_+
    &=
    Y\otimes X\, , 
    \qquad\ 
\left.
    Y_{3j-1}X_{3j}
    \right|_+
    =
    I\otimes Y\, , 
    \qquad\ \ \, 
\left.
    Y_{3j-2}Z_{3j-1}X_{3j}
    \right|_+
    =
    Y\otimes Z\, , 
    \nonumber\\
    %%%%%
            \left.
Y_{3j-2}Y_{3j-1}
    \right|_+
    &=
    Y\otimes Y\, , 
    \qquad\ \ 
\left.
    Y_{3j-1}Y_{3j}
    \right|_+
    =
    -Z\otimes X\, , 
    \qquad
\left.
    Y_{3j-2}Z_{3j-1}Y_{3j}
    \right|_+
    =
    -X\otimes I\, . 
    \nonumber\\
    %%%%%
\end{align}
These combinations resemble those in the clusters spin-SYK. Therefore, if we consider each parity sector (e.g., all-$+$ sector), the gauged clusters SYK model with $M=6$ is almost the same as the clusters spin-SYK model defined in \eqref{Hamiltonian:cluster_spin_SYK}. 
Note that the dimension of the fixed-parity-$+$ sector is $4^L$, and there are $2^L$ fixed-parity sectors.

We can also allow two fermions from the same site to interact, resulting in 
\begin{align}
    H
    =
    \sum_{i<j}
    \sum_{r,r'\in \mathcal{I}'_i}
    \sum_{s,s'\in \mathcal{I}'_j}
    J_{rr'ss'}\, 
    \chi_{r}
    \chi_{r'}
    \chi_{s}
    \chi_{s'}\, . 
\label{Hamiltonian_cluster_SYK_no_overlap}
\end{align}
where the distinction between the $\psi$ and $\chi$ fermions is summarized in \eqref{numbering_fermion} and $\mathcal{I}'_j=\{6j-5,\ldots,6j\}$, and we assume $r<r'$ and $s<s'$. This differs only slightly from the version discussed above.

If we increase the window size further (beyond 3) to be of order $N$, by taking $\mathcal{I}_j=\{k(j-1)+1,\ldots,kj\}$ and $\mathcal{I}'_j=\{2k(j-1)+1,\ldots,2kj\}$ ($k=3,4,5,\ldots$), it will be hard to tell the difference from the original SYK.
%%%%%%%%%%%%%%%%%%%%%%
%%%%%%%%%%%%%%%%%%%%%%
\subsubsection{Overlapping clusters SYK model}\label{sec:q2M2SYK}
%%%%%%%%%%%%%%%%%%%%%%
%%%%%%%%%%%%%%%%%%%%%%
Another simple modification of \eqref{Hamiltonian:naive_cluster_SYK} that can avoid too many conserved parities is obtained by allowing the clusters to overlap. We can apply the same modification to \eqref{Hamiltonian:restricted_spin_SYK_ver_1} and \eqref{Hamiltonian_cluster_SYK_no_overlap}. Specifically, we consider a modification of \eqref{Hamiltonian_cluster_SYK_no_overlap} given by
\begin{align}
    H
    =
 \sum_{r_1,s_1,r_2,s_2}
J_{r_1s_1r_2s_2}\, 
    (\chi_{r_1}
    \chi_{s_1})
    (\chi_{r_2}
    \chi_{s_2})\, ,   \label{Hamiltonian:q=2_M_overlapping_clusters_SYK}
\end{align}
where $J_{r_1s_1r_2s_2}\neq 0$ only for $r_1<s_1<r_2<s_2$, $s_1-r_1<M$, and $s_2-r_2<M$. 
We call this model the $\tilde{q}=2$ overlapping clusters SYK. Here, we consider arbitrary clusters of length $M$, where the parameter $M$ controls the intra‑cluster window of the fermion block. The clusters have overlaps, although the interactions in the Hamiltonian are only between non-overlapping pairs. This simple modification eliminates most of the conserved charges, leaving only the total spin (equivalently, the fermion-number parity). We can also have a generic $\tilde{q}$-local version, 
\begin{align}
    H
    =
 \sum_{r_1,s_1,\ldots,r_{\tilde{q}},s_{\tilde{q}}}
J_{r_1s_1\cdots r_{\tilde{q}}s_{\tilde{q}}}\, 
    (\chi_{r_1}
    \chi_{s_1})
    \cdots
    (\chi_{r_{\tilde{q}}}
    \chi_{s_{\tilde{q}}})\, , 
    \label{Hamiltonian:q_local_M_overlapping_clusters_SYK}
\end{align}
where $J_{r_1s_1\cdots r_{\tilde{q}}s_{\tilde{q}}}\neq 0$ only for $r_1<s_1<\cdots<r_{\tilde{q}}<s_{\tilde{q}}$ and  $s_i-r_i<M$ for $i=1,\ldots,{\tilde{q}}$.  By taking $M$ large, we can reproduce the original SYK model with $2\tilde{q}$-body interaction. 

As the simplest model of this kind, let us consider $\tilde{q}=2$, $M=2$ overlapping clusters SYK, whose Hamiltonian is defined by 
\begin{align}
    H
    =
 \sum_{r}
    \sum_{s>r+1}
J_{rs}\, 
    (\chi_{r}
    \chi_{r+1})
    (\chi_{s}
    \chi_{s+1})\, . 
\label{Hamiltonian:q=2_M=2_overlapping_clusters_SYK}
\end{align}
$\chi_{r}\chi_{r+1}$ becomes $\mathrm{i}Z_{j}$ (for $r=2j-1$) or $\mathrm{i}X_{j}X_{j+1}$ (for $r=2j$). 
Therefore, $(\chi_{r}\chi_{r+1})(\chi_{s}\chi_{s+1})$ can be written by using only tensor products of $X$, $Z$, and $I$, and hence, the Hamiltonian can be written as a real symmetric matrix. Therefore, if this model is chaotic, then we expect the GOE universality class. See the numerical analysis in Sec.~\ref{sec:Numerics-SYK} which demonstrates the GOE universality class. 

In the original SYK model, the particle-hole symmetry (see e.g., Appendix A of Ref.~\cite{Cotler:2016fpe}) leads to a two-fold degeneracy of the energy eigenvalues for $N\ \mathrm{mod}\ 8=2,4,6$. When $N\ \mathrm{mod}\ 8=4$, there is a two-fold degeneracy in each parity sector. When $N\ \mathrm{mod}\ 8=2,6$, different parity sectors share the same spectrum. When $N\ \mathrm{mod}\ 8=0$, the particle-hole symmetry does not constrain the energy spectrum. 

The situation is different in the $q=2$, $M=2$ overlapping clusters SYK model and there is a nontrivial consequence of the particle-hole symmetry even for $N\ \mathrm{mod}\ 8=0$ because the Hamiltonian is real. Indeed, for $N\ \mathrm{mod}\ 8=0$, there is an operator $P$ corresponding to the particle-hole symmetry that commutes with the Hamiltonian $H$ and parity operator $Z\otimes\cdots\otimes Z$ :
\begin{align}
    P = Y\otimes X\otimes Y\otimes X\otimes\cdots\otimes Y\otimes X\, . 
\end{align}
Note that $P$ is an ordinary unitary operator although the particle-hole transformation in the original SYK model is anti-unitary; this is because the Hamiltonian can be written without using a complex number. Therefore, $P$, the Hamiltonian, and the parity operator can be diagonalized simultaneously, and we need to look at $P=+1$ sector and $P=-1$ sector separately, by using projected Hamiltonian $H_{\pm}=\frac{I\pm P}{2}H\frac{I\pm P}{2}$. Note that both $P$ and $H_\pm$  are real symmetric in the Jordan-Wigner basis and hence we expect the GOE statistics for nonzero eigenvalues. 

$P$ commutes with $H$ for other values of $N$, too. When $N\ \mathrm{mod}\ 8=2$ or $6$, it takes the form of $P=Y\otimes X\otimes\cdots\otimes Y$, which anticommutes with the parity $Z\otimes\cdots\otimes Z$. This leads to the degeneracy of the energy spectrum in two parity sectors. For this case, the Hamiltonian restricted to each parity sector is real symmetric, and we expect the GOE statistics. When $N\ \mathrm{mod}\ 8=4$, $P$ is pure imaginary and antisymmetric, and it commutes with the parity operator.  Because $H$ is real symmetric, the energy eigenstates can be chosen to be real vectors $\vec{v}$ in the Hilbert space. Then, $\vec{v}$  and $P\vec{v}$ are orthogonal to each other because $P$ is antisymmetric. This leads to two-fold degeneracy of the energy spectrum in each parity sector. For this case, $H_{\pm}$ is not real symmetric, and we expect the GUE universality class.
Note that the present case, with interactions limited to those between two odd-numbered and two even-numbered fermions, can be interpreted as a two-local version of the bipartite SYK model studied in \cite{Fremling_2022}, where the shift in the symmetry class above has also been discussed.
%%%%%%%%%%%%%%%%%%%%%%
%%%%%%%%%%%%%%%%%%%%%%
\subsection{Toward quantum simulations}\label{sec:overlapping_SYK_quantum_simulation}
%%%%%%%%%%%%%%%%%%%%%%
%%%%%%%%%%%%%%%%%%%%%%
One of the motivations for this paper was to find a model more easily simulatable on quantum computers. Specifically, we are interested in near-term devices which do not have perfect quantum error corrections. For concreteness, let us consider Hamiltonian time evolution via Suzuki-Trotter decomposition, and focus on the counting of two-qubit gates, specifically, controlled NOT (CNOT) gates for two main reasons. First, CNOT gates are the only 2-qubit gates required in a universal gate set to simulate an arbitrary unitary and, second, such gates are typically the most costly to implement.
We use $\textrm{CNOT}_{p,q}$ to denote the CNOT gate which uses the qubit $p$ as the control qubit and the qubit $q$ as the target qubit:
\begin{align}
    &
    \textrm{CNOT}_{p,q}(\ket{0}_p\ket{0}_q)
    =
    \ket{0}_p\ket{0}_q\, , 
    \qquad
    \textrm{CNOT}_{p,q}(\ket{0}_p\ket{1}_q)
    =
    \ket{0}_p\ket{1}_q\, ,\nonumber\\ 
    &   
    \textrm{CNOT}_{p,q}(\ket{1}_p\ket{0}_q)
    =
    \ket{1}_p\ket{1}_q\, , 
    \qquad
    \textrm{CNOT}_{p,q}(\ket{1}_p\ket{1}_q)
    =
    \ket{1}_p\ket{0}_q\, . 
\end{align}
The Hamiltonian of the SYK model can also be written as a sum of Pauli strings. For the clusters SYK model with cluster size $M$, each cluster is a Pauli string with at most length $\lceil\frac{M}{2}+1\rceil$, and hence a $q$-local Hamiltonian consists of Pauli strings with at most length $q\lceil\frac{M}{2}+1\rceil$. The unitary time evolution is approximated by a product of the terms of the form $\exp\left(-\mathrm{i}\varepsilon\sigma\cdots\sigma\right)$. By applying one-qubit gate, we can make all Pauli operators to $\sigma_3=Z$, as $
\sigma_1 =X= hZh$ and $\sigma_2=Y=sXs^\dagger = shZ(sh)^\dagger$, where $h$ and $s$ are Hadamard gate and phase gate defined by 
\begin{align}
    h
    =
    \frac{1}{\sqrt{2}}
    \left(
        \begin{array}{cc}
            1 & 1\\
            1 & -1
        \end{array}
    \right)\, , 
    \qquad
    s
    =
    \left(
        \begin{array}{cc}
            1 & 0\\
            0 & \mathrm{i}
        \end{array}
    \right)\, .      
\end{align} 
Therefore, up to the conjugation by these one-qubit gates, we need to consider a unitary of the form $\exp\left(-\mathrm{i}\varepsilon\sigma_{z,p_1}\cdots\sigma_{z,p_\ell}\right)$, where $\ell$ is at most $q\lceil\frac{M}{2}+1\rceil$. 
By using 
\begin{align}
Z_{p}Z_{q}
=
\textrm{CNOT}_{p,q}
Z_{q}
\textrm{CNOT}_{p,q}\, , 
\label{eq:Z-and-CNOT}
\end{align}
we can rewrite it with one-qubit rotation and $2(\ell-1)$ CNOT gates:
\begin{align}
&
\exp\left(-\textrm{i}\varepsilon Z_{p_1}\cdots Z_{p_\ell}\right)
\nonumber\\
&\quad=
\textrm{CNOT}_{p_1,p_2}
\cdots
\textrm{CNOT}_{p_{\ell-1},p_\ell}
\exp\left(-\textrm{i}\varepsilon Z_{p_\ell}\right)
\textrm{CNOT}_{p_{\ell-1},p_\ell}
\cdots
\textrm{CNOT}_{p_1,p_2}\, . 
\end{align}
In the original SYK model, the typical length of Pauli strings and the number of CNOT gates are proportional to the number of fermions $N$ and the number of terms in the Hamiltonian scales as $N^4$ (or $N^q$ for SYK$_q$). In the (overlapping) clusters SYK models, the reduction in CNOT gate cost is three-fold. First, as explained above, the Pauli strings are upper bounded by $\lceil\frac{M}{2}+1\rceil$ so if we take $M$ substantially smaller than $N$, the number of CNOT gates can be reduced significantly. Second, in the case where $M$ is in again substantially smaller than $N$, the total number of terms in the Hamiltonian is reduced as well, and is of order $\mathcal{O}(L^2M^4)$, where we recall that $L=N/4$.
This reduces the number of exponentials per Trotter slice and as such reduces the CNOT gate cost.
A final reduction in gate cost arises in quantum devices with architectures where Hamiltonian terms involving qubits or qudits having nearby indices are less costly than those involving more distant ones. For example, if the qubits are ions, qubits sitting nearby could be moved together more easily than distant ones.
We note that, while the first reduction in gate cost does not apply to the spin-SYK models, the other two do.
In the following sections, we will study which features of the SYK model are preserved in the qudit models and (overlapping) clusters SYK model for $M=2$.
The clusters SYK model can approach the original SYK systematically, by making $M$ bigger. As mentioned in the introduction, it is important to check which features of the SYK model are preserved in the clusters SYK. 

Complementing the gate‑count perspective, multiple platforms already realize the two-qubits-per-site structure we target. In photonics, single‑photon ququarts have been prepared and characterized using polarization–orbital‑angular‑momentum encoding \cite{nagali2010experimental}.
Building on that encoding, two-ququart entanglement has been demonstrated with linear optics using two independent ququarts \cite{kwon2014entangling} and exploited for coding with high-fidelity four-dimensional Bell states \cite{hu2018beating}. Beyond optics, trapped-ion processors now support native two-qudit entangling gates up to dimension five, which directly include ququart--ququart entanglers \cite{hrmo2206native}. Finally, neutral-atom architectures have proposed concrete schemes that encode two qubits in a single ytterbium-171 atom and compile inter-ququart Rydberg gates together with intra-ququart operations, offering a clear path to experimental realization \cite{jia2024architecture}. These results indicate that the two-local qudit and cluster models studied here align with near-term capabilities.

Beyond gate-count considerations, several architectures already support the primitives our models require.
Trapped‑ion processors realize universal single-qudit logic and demonstrate genuine multilevel entangling gates, with experiments operating at local dimensions up to seven \cite{Ringbauer:2021lhi, hrmo2206native}. Superconducting circuits now implement high‑fidelity two‑qutrit controlled‑phase gates, validated by randomized benchmarking and cross‑entropy metrics \cite{goss2022high}. Optically addressable ytterbium‑ion registers combine single‑qudit rotations with a Mølmer–Sørensen interaction in a two‑ququart device \cite{aksenov2023realizing}.

%%%%%%%%%%%%%%%%%%%%%%
%%%%%%%%%%%%%%%%%%%%%%
\section{Numerical analyses}\label{sec:Numerics}
%%%%%%%%%%%%%%%%%%%%%%
%%%%%%%%%%%%%%%%%%%%%%

%%%%%%%%%%%%%%%%%%%%%%
%%%%%%%%%%%%%%%%%%%%%%
\subsection{Numerical analyses of qudit models}\label{sec:Numerics-qudit}
%%%%%%%%%%%%%%%%%%%%%%
%%%%%%%%%%%%%%%%%%%%%%
First, we study the qudit SYK model defined by \eqref{Hamiltonian:Qudit_SYK} for $d=3,4,5,6$ and $q=2$, and $d=3$ and $q=3$. The data presented below are consistent with quantum chaos, except for a small fraction of energy spectrum near the edges.

We normalize the Hamiltonian so that the variance of the energy is 1, or equivalently, $\langle\mathrm{Tr} H^2\rangle=d^L$. 
Because
\begin{align}
    \langle\mathrm{Tr} H^2\rangle
    &=
    \sum_{i_1<i_2<\cdots<i_q}
    \sum_{\alpha_1,\ldots,\alpha_q}
    \langle J_{i_1,\alpha_1;\cdots;i_q,\alpha_q}^2\rangle
    \mathrm{Tr} (T_{i_1,\alpha_1}^2\cdots T_{i_q,\alpha_q}^2)\\
    &= J^2\binom{L}{q}(d^2-1)^q 2^qd^{L-q}\, , 
\end{align}
we take $J$ such that $\langle\mathrm{Tr} H^2\rangle$ becomes $d^L$:
\begin{align}
    J^2 = \left[\binom{L}{q}\left(\frac{2(d^2-1)}{d}\right)^q \right]^{-1}\, .\label{eqn:SUqSpin-normalization}
\end{align}

\begin{figure}
    \includegraphics[]{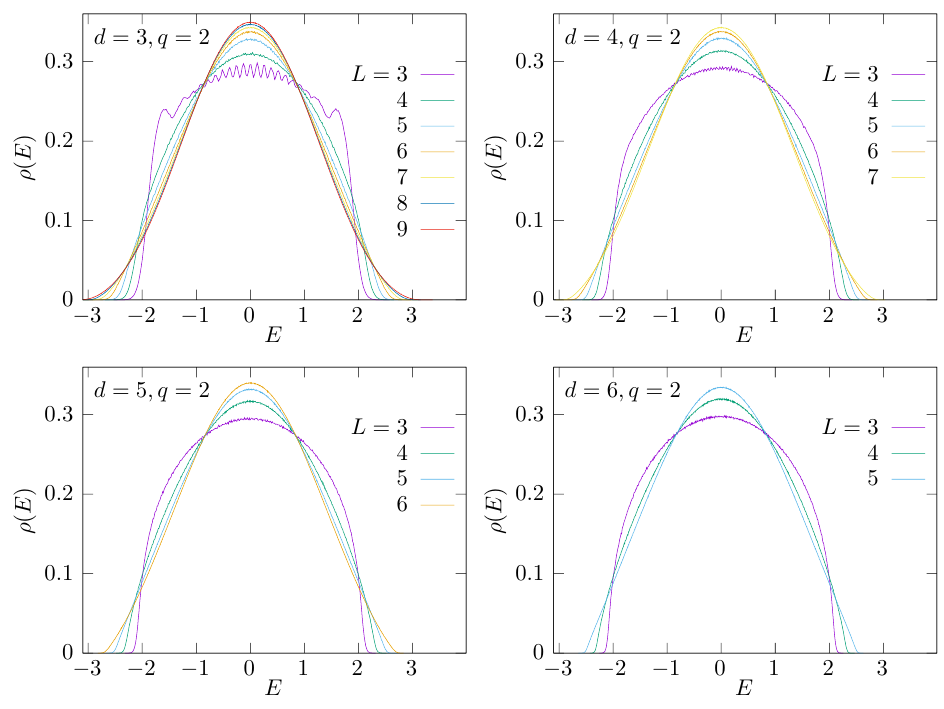}
    \includegraphics[]{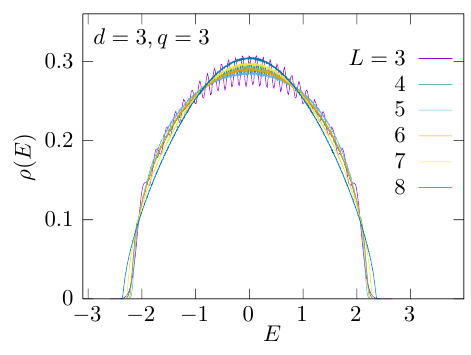}
    \caption{
Plots of the density of states $\rho(E)$ for $d>2$ qudit SYK model \eqref{Hamiltonian:Qudit_SYK} with the coupling constant normalization \eqref{eqn:SUqSpin-normalization}.
The $q=2$ model shows soft edges, while $q=3$ shows hard edges.     
The number of samples is $2^5\times 3^{12-L}$ for $d=3$ (for both $q=2$ and $q=3$), $5\times 4^{11-L}$ for $d=4$, $2^8\times 5^{7-L}$ for $d=5$, $2\times 6^{9-L}$ for $d=6$, so that the number of eigenvalues is at least $1.7\times10^7$.}
    \label{fig:DoS_qudit}
\end{figure}

Practically, we compute all energy eigenvalues ordered as $E_0\le E_1\le E_2\le\cdots$ by diagonalizing the Hamiltonian. We did not see degeneracy of energy levels in the qudit SYK.  
%%%%%%%%%%%%%%%%%%%%%%%%%%%%%%%%%%%%%%%%%%%%%%%
\subsubsection*{Density of states}
%%%%%%%%%%%%%%%%%%%%%%%%%%%%%%%%%%%%%%%%%%%%%%%
The density of states is shown in Fig.~\ref{fig:DoS_qudit}. 
We see that $q=2$ model shows soft Gaussian edges, while $q=3$ shows hard square-root-like edges. Note that, even for $q=2$, the edge becomes harder as $d$ is increased while $L$ is fixed.
On the other hand, fixing $q=2$ and $d$, an increase of $L$ should also soften the edges.
The soft edge behavior, which appears to be related to the values of the neighboring gap ratio at low $i$ being smaller than the RMT values (see Figure \ref{fig:gapratio_each_i}), could indicate that the model becomes less chaotic at low temperatures.
Similar observations were made in the context of related models, including the sparse SYK model and local SYK models, see, \textit{e.g.}, \cite{Garcia-Garcia:2020cdo,Anegawa_2023,Orman:2024mpw,Altland:2024ubs}.
As explained in \cite{Garcia-Garcia:2020cdo,Altland:2024ubs}, the soft edge could be a result of the (lack of suppression) of collective spectral fluctuations, including ``breathing modes'' of spectrum which change the width per sample.
The suppression of these fluctuations in disordered-averaged models, like SYK, goes as $1/K$ with $K$ the number of random parameters in the model.
In other words, the less random the model, the less chaotic and the softer the edge of the density of states.
This could explain our observations of the parameter dependence of the density of states, where for example the $q=2$ qudit model has considerably less random parameters than the $q=3$ model and one expects the collective fluctuations to become less suppressed.
Furthermore, an increase in $d$ increases the number of random parameters, while an increase in $L$ (keeping $q$ and $d$ fixed) results in a relatively sparser model than full SYK, again consistent with a (lack of) suppression of the collective fluctuations.
We expect that, to retain hard edges for large $L$, one either needs to consider $q>2$ or scale $d$ in some way with $L$.
It would be interesting to find a precise scaling and we leave this question to future work.
We will make further comments in Section \ref{sec:conclusion} and also comment on the implications for a holographic dual.

%%%%%%%%%%%%%%%%%%%%%%%%%%%%%%%%%%%%%%%%%%%%%%%
\subsubsection*{Nearest-Neighbor Level Spacings}
\begin{figure}[htbp]
	\centering
        \includegraphics[width=\textwidth]{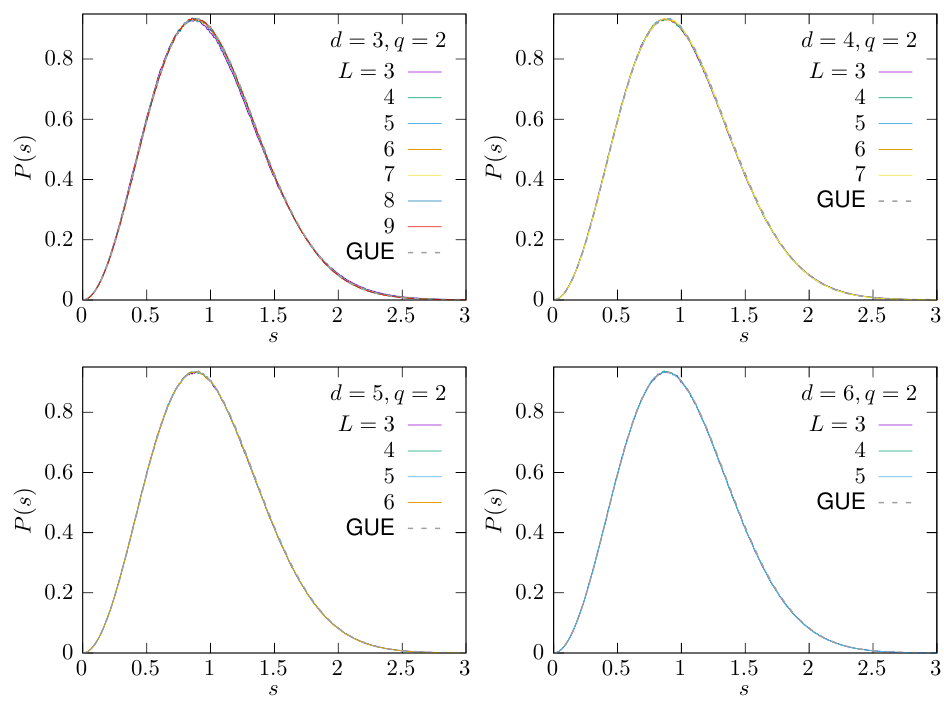}
        \includegraphics[width=0.5\textwidth]{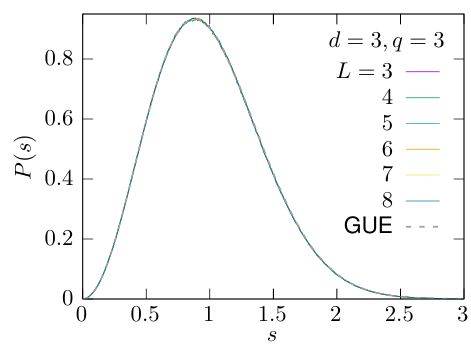}
	\caption{Nearest-Neighbor Level Spacings for qudit SYK model \eqref{Hamiltonian:Qudit_SYK} for $d=3, 4, 5, 6$ and $q=2$, and for $d=3$ and $q=3$. Multiple values of sites $L$ are plotted in the same figure. The distribution converges to that of GUE random matrix as $L$ increases.
    %The number of samples is $2^5\times 3^{12-L}$ for $d=3$, $5\times 4^{11-L}$ for $d=4$, $2^8\times 5^{7-L}$ for $d=5$, $2\times 6^{9-L}$ for $d=6$, so that the number of eigenvalues is at least $1.7\times10^7$.
    The GUE result is from \cite{Dietz:1990}.}
	\label{fig:nnls_distributions}
\end{figure}

%%%%%%%%%%%%%%%%%%%%%%%%%%%%%%%%%%%%%%%%%%%%%%%
We examined the spacing between adjacent energy levels $s_i=E_{i+1}-E_i$ in the generalized spin model.
We conducted polynomial unfolding on the eigenvalues, generating the distributions of nearest-neighbor level spacings across various spatial points $L$. Our findings indicate that the results are consistent with the expected patterns of the Wigner surmise within the GUE class. The corresponding plots for these distributions are illustrated in Fig.~\ref{fig:nnls_distributions}.

%%%%%%%%%%%%%%%%%%%%%%%%%%%%%%%%%%%%%%%%%%%%%%%
\subsubsection*{Nearest-gap ratio}
\begin{figure}
    \includegraphics[]{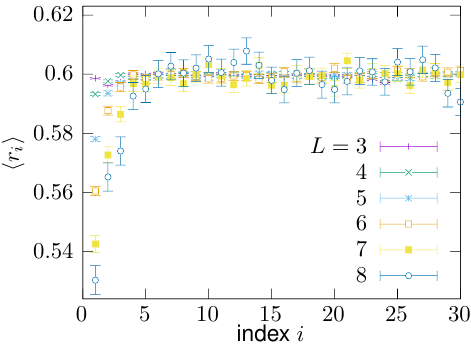}
    \includegraphics[]{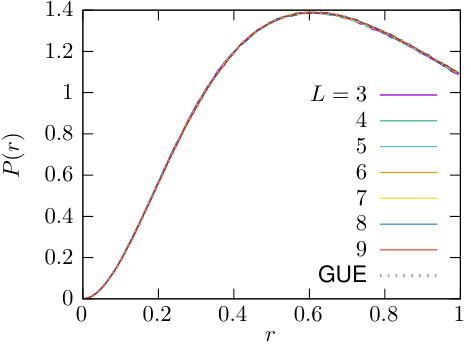}\\
    \includegraphics[]{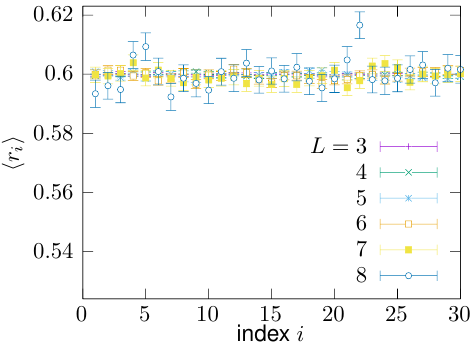}
    \includegraphics[]{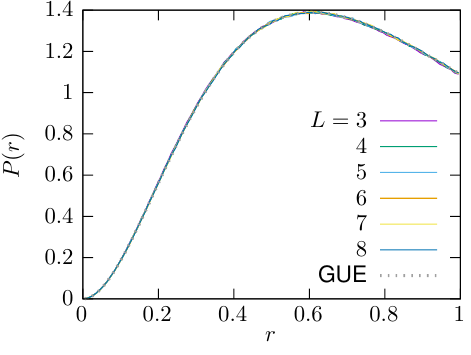}
    \caption{
    Neighboring gap ratio for the fixed-$i$-unfolded spectrum for the first $30$ neighboring gap pairs for $d=3$ qudit SYK model \eqref{Hamiltonian:Qudit_SYK} (left) and the distribution of the gap ratio over the entire energy spectrum (right). [Top] $q=2$. [Bottom] $q=3$.
    }
    \label{fig:gapratio_each_i}
\end{figure}

%%%%%%%%%%%%%%%%%%%%%%%%%%%%%%%%%%%%%%%%%%%%%%%
The nearest-gap ratio is defined by $r_i\equiv\mathrm{min}\left(
\frac{s_{i+1}}{s_{i}}, \frac{s_{i}}{s_{i+1}}
\right)$. 
We plotted the average $\langle r_i\rangle$ for each $i$, as shown in Fig.~\ref{fig:gapratio_each_i}. We can see a good agreement with the GUE value 0.59975 \cite{Nishigaki:2024} except for $q=2$ at very small $i$.
As commented on above, this appears to be related to the soft edges of the density of states and may be related to less chaotic behavior of the model at low temperatures.

%%%%%%%%%%%%%%%%%%%%%%
%%%%%%%%%%%%%%%%%%%%%%
\subsubsection*{Spectral Form Factor}
\begin{figure}
    \includegraphics[width=16.0cm]{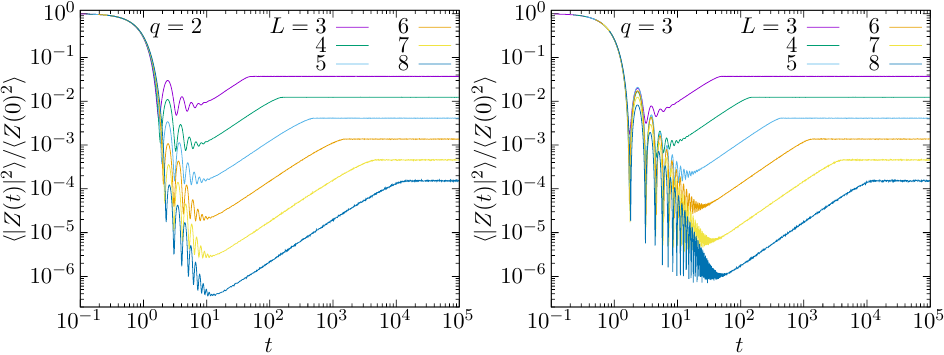}
    \caption{
    $|Z(t)|^2/|Z(0)|^2$ for various $L$ for $d=3$ qudit SYK model \eqref{Hamiltonian:Qudit_SYK}.
    Left: $q=2$. Right: $q=3$.
        The curves are in consecutive order with regards to $L$.
}\label{sec:SFF_d=3}
\end{figure}

%%%%%%%%%%%%%%%%%%%%%%
%%%%%%%%%%%%%%%%%%%%%%
The spectral form factor (SFF) can capture the correlation of energy eigenvalues in a wide range of the energy spectrum. The SFF is defined by $|Z|^2$, where $Z(t)=\sum_{j}e^{-\mathrm{i}tE_j}$. The late-time behavior of the SFFs of chaotic systems follows the universal pattern depending on their universality class. The SFF of the qudit SYK model shown in Fig.~\ref{sec:SFF_d=3} exhibits this universal pattern of the GUE universality class. We can see a clean ramp proportional to $t$, which indicates a good agreement with random matrix theory in a wide range of the energy spectrum.  

%%%%%%%%%%%%%%%%%%%%%%
%%%%%%%%%%%%%%%%%%%%%%
   \subsection{Numerical analyses of overlapping clusters SYK model}\label{sec:Numerics-SYK}
%%%%%%%%%%%%%%%%%%%%%%
%%%%%%%%%%%%%%%%%%%%%%
Next, we study the $q=2$, $M=2$ overlapping clusters SYK model (Sec.~\ref{sec:q2M2SYK}). We define the normalization of the Gaussian random coupling in such a way that the variance of the energy does not depend on the cluster size $M$, and hence, takes the same value as the original SYK model with four-local interactions. We take the variance of the random coupling in the original SYK model to be $6/N^3$.

The Hamiltonian of $q=2$, general $M$ overlapping clusters SYK is given by \eqref{Hamiltonian:q=2_M_overlapping_clusters_SYK}.
The special case $M=2$ is also given in \eqref{Hamiltonian:q=2_M=2_overlapping_clusters_SYK}.
In the case $M=2$, the number of couplings $J_{pq}$ is given by $\binom{N-2}{2}$, while the number of couplings for the original SYK model is $\binom{N}{4}$.
Therefore, the variance of the couplings, $J^2$, should be chosen as
\begin{align}
    J^2 = \frac{6}{N^3}\frac{\binom{N}{4}}{\binom{N-2}{2}}
    =\frac{N-1}{2N^2}\, .\label{eqn:q2M2normalization}
\end{align}

To see chaotic features for $M=2$, we take into account parity and particle-hole symmetry. Specifically:
\begin{enumerate}
\item
For $N\ \mathrm{mod}\ 8=2, 6$, parity even and odd sectors share the same spectrum, and hence we use only the parity even sector. 

\item
For $N\ \mathrm{mod}\ 8=4$, each parity sector has a two-fold degenerate energy spectrum. We remove the degeneracy and study the parity even and odd sectors separately. To compute the density of states, SFF and gap ratio, we take the average over the two sectors. 

\item
For $N\ \mathrm{mod}\ 8=0$, we look at the four sectors separately: even/odd under parity and particle-hole transformation. To compute the density of states, SFF and gap ratio, we take the average over four sectors. 

\end{enumerate}
If the system is chaotic, we expect the GOE universality class for $N\, \mathrm{mod}\, 8=0,2,6$ and the GUE universality class for $N\, \mathrm{mod}\, 8=4$, respectively. 

%%%%%%%%%%%%%%%%%%%%%%%%%%%%%%%%%%%%%%%%%%%%%%%
\subsubsection*{Density of states}
\begin{figure}[htbp]
\centering
    \includegraphics[width=8.0cm]{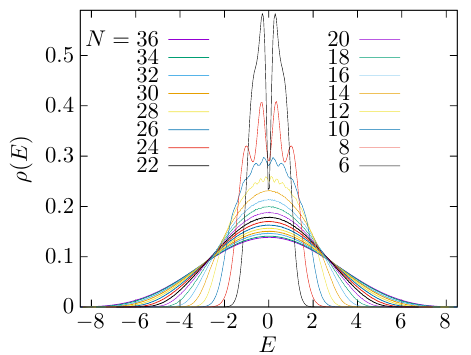}
	\caption{Density of states in $q=2$, $M=2$ overlapping clusters SYK model \eqref{Hamiltonian:q=2_M=2_overlapping_clusters_SYK}. 
    For $6\leq N\leq 34$, $2^{24-N/2}$ samples are used. For $N=36$, $11$ samples are used.
   }
\label{fig:q2M2_clusters_SYK_DoS}
\end{figure}

%%%%%%%%%%%%%%%%%%%%%%%%%%%%%%%%%%%%%%%%%%%%%%%
In Fig.~\ref{fig:q2M2_clusters_SYK_DoS}, we show the density of states. The edge looks different from the original SYK model; it is soft, unlike the hard edge of the original SYK. 
Note that we observed soft, Gaussian edges in the $q=2$ qudit SYK model, too. 
In this model, an increase of $M$ should lead to harder, or $\sqrt{E}$, edges for the same reasons mentioned in the discussion of the density of states of the qudit models.
We expect a similar parameter dependence of the spectrum as discussed there, where now the role of $d$ is played by $M$.
In particular, as noted in Section \ref{sec:q2M2SYK}, when $M$ increases to order $L$ our model becomes the full SYK model whose spectrum has hard edges.
We return to this point in Sec.~\ref{sec:conclusion}.

%%%%%%%%%%%%%%%%%%%%%%%%%%%%%%%%%%%%%%%%%%%%%%%
\subsubsection*{Nearest-Neighbor Level Spacings}
\begin{figure}[htbp]
    \includegraphics{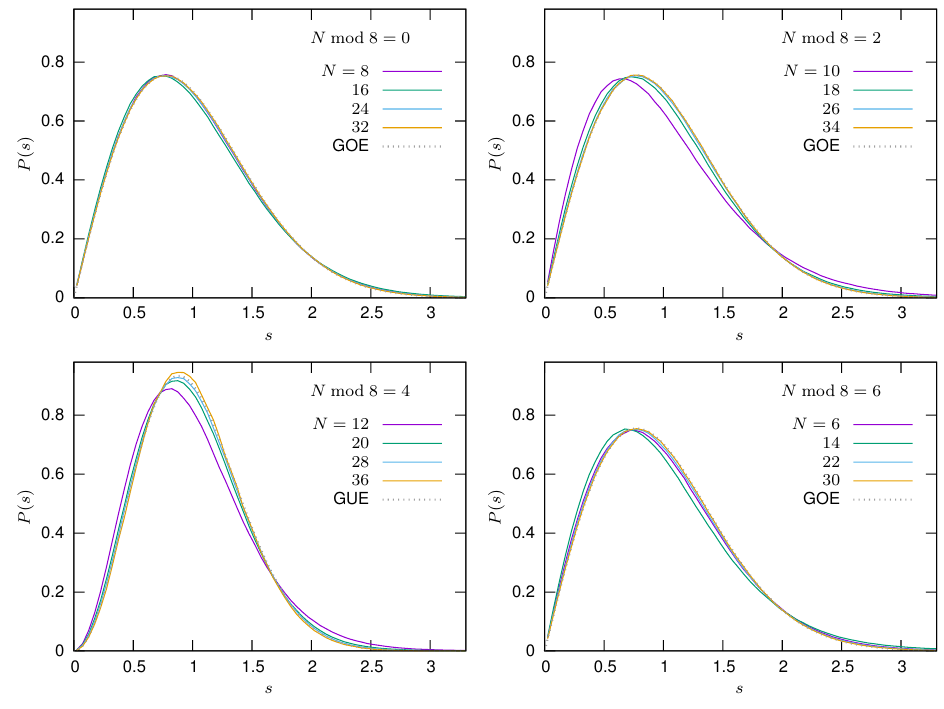}
	\caption{
    Nearest-neighbor level spacing distribution in $q=2$, $M=2$ overlapping clusters SYK model \eqref{Hamiltonian:q=2_M=2_overlapping_clusters_SYK}. 
    }
\label{fig:q2M2_clusters_SYK_Wigner_surmise}
\end{figure}

%%%%%%%%%%%%%%%%%%%%%%%%%%%%%%%%%%%%%%%%%%%%%%%
In Fig.~\ref{fig:q2M2_clusters_SYK_Wigner_surmise}, the nearest-neighbor level spacing is plotted. We can see a reasonably good agreement with the GOE universality class for $N\, \mathrm{mod}\, 8=0,2,6$ and the GUE universality class for $N\, \mathrm{mod}\, 8=4$, which becomes better at larger $N$. 

%%%%%%%%%%%%%%%%%%%%%%%%%%%%%%%%%%%%%%%%%%%%%%%
\subsubsection*{Nearest-gap ratio}
\begin{figure}[htbp]
    \centering
    \includegraphics{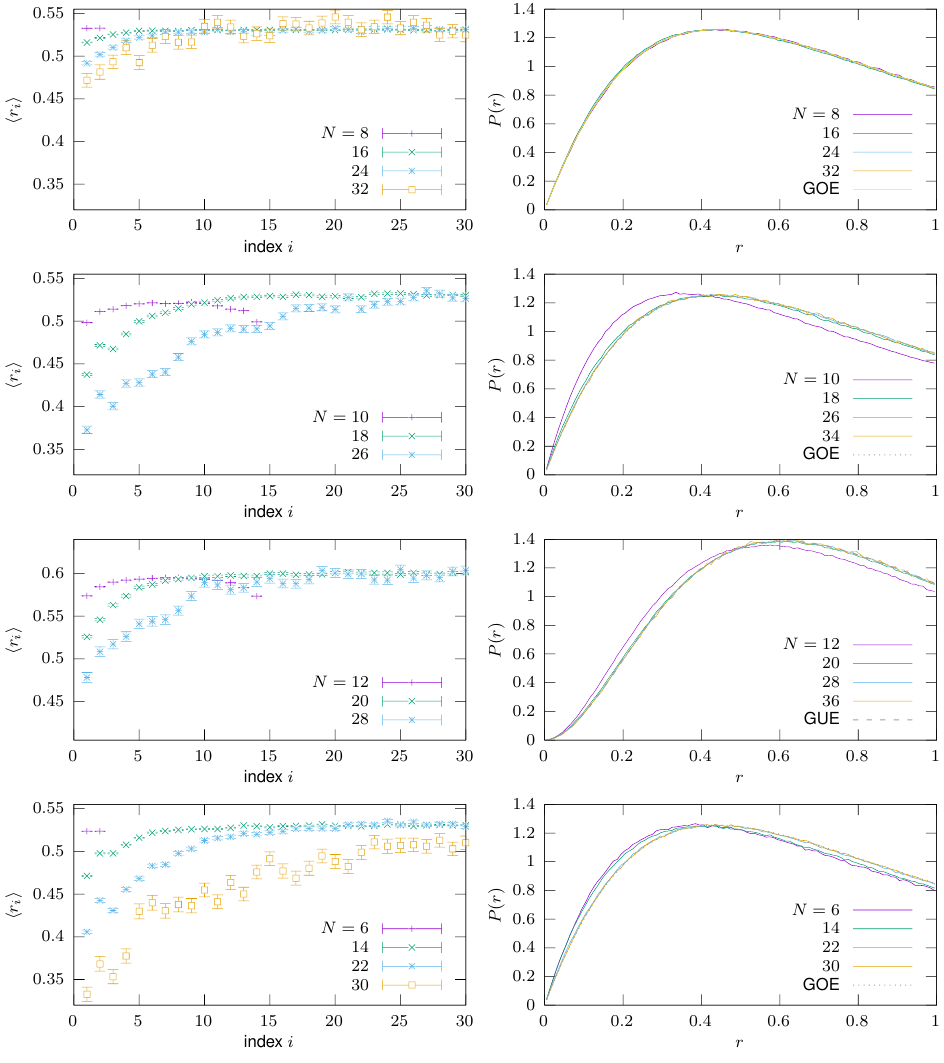}
    \caption{$q=2$, $M=2$ overlapping clusters SYK model \eqref{Hamiltonian:q=2_M=2_overlapping_clusters_SYK}, nearest-neighbor gap ratio for the fixed-$i$-unfolded spectrum. 
    For $N\ \mathrm{mod}\ 8=0$, the average over four sectors (even/odd under parity and particle-hole transformation) is taken.
    For $N\ \mathrm{mod}\ 8=2,6$, the spectrum is identical between the parity even and odd sectors. 
    For $N\ \mathrm{mod}\ 8=4$, the average over parity even and odd sectors is taken. 
    }
\label{fig:q2M2_clusters_SYK_r}
\end{figure}

%%%%%%%%%%%%%%%%%%%%%%%%%%%%%%%%%%%%%%%%%%%%%%%
In Fig.~\ref{fig:q2M2_clusters_SYK_r}, the nearest-gap ratio is plotted. Again, we observed reasonably good agreement with GOE universality class for $N\, \mathrm{mod}\, 8=0,2,6$ and the GUE universality class for $N\, \mathrm{mod}\, 8=4$, except for a small fraction of the spectrum close to the edges. 
We note that this fraction increases with increasing $N$.
This feature is related to the soft edge of the density of states observed above.
Similarly to the density of states, this feature disappears when $M$ is increased.
We further discuss this point in Sec.~\ref{sec:conclusion}.

%%%%%%%%%%%%%%%%%%%%%%
%%%%%%%%%%%%%%%%%%%%%%
\subsubsection*{Spectral Form Factor}
\begin{figure}[htbp]
    \centering
    \includegraphics{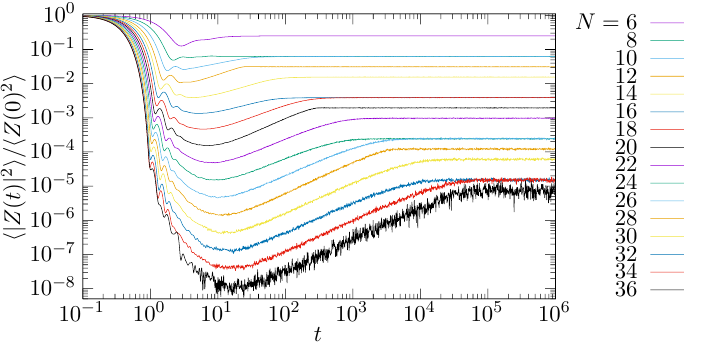}
	\caption{$q=2$, $M=2$ overlapping clusters SYK model \eqref{Hamiltonian:q=2_M=2_overlapping_clusters_SYK}, Spectral Form Factor. 
    For $N\ \mathrm{mod}\ 8=0$, the average over four sectors (even/odd under parity and particle-hole transformation) is taken.
    For $6\leq N\leq 34$, $2^{24-N/2}$ samples are used. For $N=36$, $11$ samples are used.
        The curves are in consecutive order with regards to $N$.
}
\label{fig:q2M2_clusters_SYK_SFF}
\end{figure}

%%%%%%%%%%%%%%%%%%%%%%
%%%%%%%%%%%%%%%%%%%%%%
In Fig.~\ref{fig:q2M2_clusters_SYK_SFF}, the spectral form factor is plotted, where the sum over the eigenvalues is taken over all the symmetry sectors. For $N\ \mathrm{mod}\ 8=0$ there is no degeneracy and for $N\ \mathrm{mod}\ 8=2,4,6$ there is a twofold degeneracy, so the late-time average is $2^{-N/2}$ for $N\ \mathrm{mod}\ 8=0$ and $2^{1-N/2}$ otherwise. We can see a clean ramp emerges, suggesting the spectral rigidity over a wide range in the energy spectrum. 

%%%%%%%%%%%%%%%%%%%%%%
%%%%%%%%%%%%%%%%%%%%%%
\section{Conclusions and discussion}\label{sec:conclusion}
%%%%%%%%%%%%%%%%%%%%%%
%%%%%%%%%%%%%%%%%%%%%%
In this paper, we considered the qudit SYK model (Sec.~\ref{sec:qudit_model_definition} and Sec.~\ref{sec:Numerics-qudit}), clusters spin-SYK model (Sec.~\ref{sec:clusters-spin-SYK}), and clusters SYK model (Sec.~\ref{sec:clusters-SYK} and Sec.~\ref{sec:Numerics-SYK}).
These models contain the original SYK model as a special case. 

It is natural to expect that the $q$-local qudit SYK model exhibits stronger chaotic behavior than the original SYK model with the $q$-local interaction. Specifically, it would be interesting to see if the Maldacena-Shenker-Stanford (MSS) bound~\cite{Maldacena:2015waa} is saturated. Even for $q=2$, the MSS bound might be saturated in the large-$d$ limit. This is because the large-$d$ limit of the two-local qudit SYK model resembles the large-$M$ limit of two-local clusters spin-SYK and SYK models, where $M$ is the cluster size, which is close to the original SYK model with four-local interaction. 

For the $q$-local clusters SYK and spin-SYK, we can increase the cluster size $M$ and see the convergence to the $2q$-local original SYK model.
This is particularly interesting for $q=2$. If some features of the original $2q=4$ SYK model can be captured by the $q=2$ clusters SYK model at small values of $M$, and if small $M$ and large $M$ regions are smoothly connected, then we could use the small-$M$ region as a model of quantum gravity to be simulated on quantum devices.  

Taking the cluster size $M$ larger in the $q=2$ clusters SYK and spin-SYK models can be seen as adding more `internal' degrees of freedom to each cluster. For instance, $\chi_\alpha\chi_\beta$ can be interpreted as an $M\times M$ matrix.
Going to larger $d$ in the qudit model can have the same interpretation. It would be interesting if $M$ and $d$ have something to do with the matrix size $N$ in gauge theory. 

An alternative simplification of the SYK model, which also makes it more amenable to quantum simulation, is to consider only a single cluster of fermions but allow more than two fermions in the cluster.
For fixed $M$, such a model effectively turns the original SYK model into a finite range, or local, SYK model (see for example \cite{Garcia-Garcia:2018pwt,Anegawa_2023,Anegawa:2023tgp}).
In these models, the random $q$-body interactions of the original SYK model are set to zero when they involve fermions which are further apart than a maximum range $M$. 
Similar to the clusters models, the resulting Hamiltonian only involves short Pauli strings, substantially decreasing the gate cost for quantum simulation.
This class of models will be studied future work.

We end this section with some comments on the soft behavior of the spectral edge in some of the models studied.
For quantum systems admitting a holographic dual, it is important that chaotic features persist to low temperatures or, equivalently, the spectral edge (see, \textit{e.g.}, \cite{Garcia-Garcia:2016mno,Garcia-Garcia:2017pzl,Garcia-Garcia:2020cdo} for discussions in the context of the (sparse) SYK model).
For example, the traversable wormhole protocol tested in \cite{Jafferis:2022crx} requires scrambling dynamics at sufficiently low temperatures \cite{Gao:2019nyj,Brown:2019hmk,Nezami:2021yaq,Schuster:2021uvg}.
As recently discussed in \cite{Altland:2024ubs}, there is a relation between the hardness of the spectral edge and the number of random parameters in the system.
For random matrix models and JT gravity, the number of parameters is polynomial in the ``Hilbert space'' dimension $D$ (\textit{i.e.}, size of the matrix) while the SYK model only has $\mathcal{O}(\log D)$ random couplings.
This distinction manifests itself, at finite $D$, in a relatively soft edge of the density of states in the case of the SYK model.
Another quantity which appears sensitive to the softness of the spectral edge is the nearest-gap ratio $\langle r_i\rangle$ for low values of $i$.

Given that some of the models defined in this work can be viewed as sparsifications of the original SYK model, we should expect even softer behavior of the spectral edge. 
This is indeed our observation in the numerical analyses of the $q=2$, $M=2$ overlapping clusters SYK model, where we observe soft tails and significant deviation of the random matrix value of $\langle r_i\rangle$ at low values of $i$.
We also observed that the deviations become more pronounced when the number of fermions $N$ increases.
A natural way to suppress these deviations in the overlapping clusters models would be to scale $M$ in some way with $N$. 
It would be interesting to establish a precise criterion for the spectral edge to be sufficiently hard, such that the model exhibits gravitational physics at low temperatures. 
For example, such a criterion may require only a small number of $\langle r_i\rangle$ to be below the RMT value. 
Given such a criterion, it would be interesting to derive a precise scaling of $M$ with $N$.

Data supporting the findings in this manuscript are available in \cite{hanada_2025_15591043}.

%%%%%%%%%%%%%%%%%%%%%%%
%%%%%%%%%%%%%%%%%%%%%%%
\section*{Acknowledgments}
%%%%%%%%%%%%%%%%%%%%%%%
%%%%%%%%%%%%%%%%%%%%%%%
We thank Damian Galante, Jad Halimeh, Philipp Hauke, Brian Swingle, and David Vegh for discussion.
SvL also thanks Anosh Joseph, Kayleigh Mathieson and Pratik Roy for collaboration on a related project.
M.~T. also thanks Indrasen Ghosh and Yoshifumi Nakata for collaboration on a related project.
A part of the computations in this work has been done using the facilities of the Supercomputer Center, the Institute for Solid State Physics, the University of Tokyo.
M.~H. thanks the STFC for the support through the consolidated Grant No. ST/Z001072/1.
M.~T. was partially supported by the Japan Society for the Promotion of Science (JSPS) Grants-in-Aid for Scientific Research (KAKENHI) Grants No. JP20K03787, JP21H05185, and JP25K00925.
O.~O. was supported by the TUBITAK Grant No. 123F353.
SvL is in part supported by the Wits-IBM Quantum Computing Seed Funding Programme, Grant No. QCSeed003/2023.
The work by M.~T. and M.~H. was partially supported by JST CREST (Grant No. JPMJCR24I2).

\bibliographystyle{apsrev4-2}
\bibliography{reference}

\end{document}